# Continuous-wave cryogenic optical absorption spectrometer for sub-THz frequencies

L. Rogić[1+], N. Somun[1+], S. Griffitt[1,2,#], A. Najev[1,†], M. Spaić[1], S. Hameed[2,‡], Y. Shemerliuk[3], S. Aswartham[3], M. Orlita[4], A. Alfonsov[3], D. Pelc[1,*]

[1]Department of Physics, Faculty of Science, University of Zagreb, Bijenička 32, HR-10000 Zagreb, Croatia
[2]School of Physics and Astronomy, University of Minnesota, Minneapolis, MN-55455 USA
[3]Leibniz Institute for Solid State and Materials Research, Helmholtzstr. 20, D-01069 Dresden, Germany
[4]Laboratoire Nationale de Champs Magnetiques Intenses, 38000 Grenoble, France

[+]equal contribution
[#]present address: Cornell University, Ithaca, USA
[†]present address: Ericsson Nikola Tesla, Zagreb, Croatia
[‡]present address: Max-Planck Institute for Solid State Research, Stuttgart, Germany
[*]correspondence to dpelc@phy.hr

**Abstract**

We present the design of a continuous-wave, highly sensitive optical spectrometer for millimeter-wave frequencies between 50 and 1000 GHz. The spectrometer uses photomixing of near-infrared light to generate radiation in a wide frequency range, and the absorbed optical power is determined directly through measurements of the sample temperature. This enables relative sensitivities of up to $10^6$ for the sample absorption coefficient below liquid-helium temperatures, suitable for measurements on highly reflective samples. The instrument is also compatible with high magnetic fields. In order to validate its performance, we measure the ferromagnetic resonance in the Mott insulator $YTiO_3$, the electron spin resonance in a standard free-radical reference compound, and the antiferromagnetic resonance in a van der Waals magnetic material.

## I. Introduction

Far-infrared spectroscopy is widely used to investigate the properties of condensed matter. The method can provide comprehensive insight into lattice dynamics, metallic conductivity and relaxation, and into collective phenomena such as charge density waves and superconductivity. However, continuous-wave optical spectroscopy becomes difficult at frequencies below about 1 THz, due to severely limited intensities of conventional (thermal) light sources. With heavy filtering and special probe/sample designs, it has been possible to extend this range down to about 250 GHz [1,2], and modern Fourier-transform infrared spectrometers can in principle operate at these frequencies. Yet most of the currently used optical instruments employ pulsed, time-domain techniques to achieve frequencies below 1 THz [3]. These devices use nonlinear elements, most commonly photoconductive antennas, to both generate broadband pulses of sub-THz radiation from ultrafast laser pulses and detect the light reflected from the sample. Advances in this technique have enabled important insights [4], including studies of soft phonons in quantum paraelectrics [5,6], spin excitations in quantum magnets [7], superconducting fluctuations [8], cyclotron resonance [9] in cuprate superconductors, etc. Nevertheless, the spectral and amplitude resolution in pulsed experiments is still somewhat limited, and given that measurements can only be performed in reflection (or transmission), it is hard to accurately determine the optical absorption in materials with reflection coefficients close to one. This large group of materials, however, includes most bulk metals and superconductors at low temperatures. Importantly, measurements on highly reflective samples are difficult with both time-domain and frequency-domain instruments. It is therefore of high interest to develop a sub-THz spectrometer that overcomes the limitations of the existing techniques.

Here we leverage recent progress in photoconductive diode technology to obtain a broadband continuous-wave source of sub-THz light [10,11,12], and combine this with a direct thermal method for the measurement of absorbed power [13,14] to construct an optical spectrometer with unprecedented sensitivity at frequencies ranging from about 50 GHz to 1 THz. The spectrometer is designed for sample temperatures below 10 K, and it is compatible with both $^4$He and $^3$He direct-access evaporation refrigerators, as well as high magnetic fields. Its unique properties enable a wide range of optical absorption measurements, and it is especially suited for highly reflective samples. The most important possible applications include measurements of metallic optical conductivity, superconducting gaps and collective modes. The instrument can also be used to study soft phonons, low-lying spin excitations and electron spin resonance (ESR) in both metals and insulators. Notably, conventional high-frequency ESR spectrometers can operate at similar frequencies, but in most cases use transmission lines and microwave technology, and therefore do not allow for frequency scans in a wide range. Instead, the frequency is kept constant, while the magnetic field is swept through the resonance [15,16,17]. Frequency-domain ESR has been reported previously, but it is much less common that the field-sweep technique and uses an array of limited-bandwidth sources to cover a wide frequency range [18,19]. Our optical setup enables continuous sweeps from 50 GHz to nearly 1 THz, along with an enhanced sensitivity due to direct thermal detection, and it is thus complementary to such instruments. In order to test the operation of the spectrometer, we perform frequency-swept ferromagnetic resonance (FMR) measurements in single crystals of the Mott insulator YTiO$_3$, ESR in the standard reference compound 2,2-diphenyl-1-picrylhydrazyl (DPPH), and antiferromagnetic resonance (AFMR) experiments in the van der Waals magnetic material Mn$_2$P$_2$S$_6$.

## II.    Instrument design
**Radiation source**
The technique used for our broadband radiation source is based on photomixing of laser light, and known from previous work [10,11,12]. Briefly, the source operates by focusing two laser beams of similar wavelengths on a photomixer diode, that then generates radiation at a frequency that corresponds to the difference of the input laser frequencies. The wavelengths of the laser light can be finely and reproducibly tuned in distributed feedback (DFB) laser diodes. Through precise control of the diode temperature, the wavelength can be changed by up to a few nanometers, which makes a wide range of sub-THz frequencies easily attainable. Efficient continuous-wave photomixers have only become commercially available within the past few years and have, to our knowledge, not been used for frequency-domain cryogenic measurements, with the exception of a very recently developed magneto-optical Kerr effect measurement setup [20].

The source of radiation in our setup consists of two DFB lasers at around 1540 nm (Eblana Photonics), with associated control units (Thorlabs) for precise diode temperature regulation; an optical coupling element (2x2 fiber coupler, Thorlabs) which combines two laser beams into one; fast photodiodes for control measurements; and a commercial fiber-integrated InGaAs photomixer (Toptica Photonics). The near-infrared part of the system is connected through optical fibers in its entirety, which eliminates alignment issues and greatly facilitates operation. The diode temperatures can be regulated in the range from zero to 50 °C. For an ambient-temperature wavelength difference of 2 nm and a wavelength temperature coefficient of 0.1 nm/K, this gives a photomixer frequency range of approximately $0 - 850$ GHz. Due to the dimensions of the antenna and the lens of the photomixer, the output power of the source is significantly reduced below ~50 GHz, which gives a practical lower frequency limit for the application of such a source. For an optical power of 30 mW at the photomixer, the maximum emitted power is greater than 200 $\mu$W at 100 GHz and approximately 10 $\mu$W at 500 GHz [12]. The frequency dependence of emitted power broadly follows the behavior of a first-order low-pass filter, $P_{gen} \sim 1/(1 + \nu^2/\nu_0^2)$, where $\nu$ is the radiation frequency, and $\nu_0 \approx 130$ GHz is the cutoff frequency [11,12]. However, the frequency dependence is strongly non-monotonic due to standing wave effects in the photomixer assembly, and the source thus requires a careful calibration. The radiation is chopped periodically by modulating the laser power to enable lock-in detection of the absorbed power [21] and to eliminate thermal drift effects.

We do not measure the laser wavelengths directly, but precisely determine the temperatures where the wavelength difference is zero, and rely on known temperature coefficients of the wavelengths to calculate the wavelength difference for any combination of diode temperatures. The zero-difference point is determined by observing beats in the combined laser signal with a fast photodiode connected to the second output port of the 2x2 fiber coupler. This procedure gives an absolute precision better than 1 GHz for the zero-difference point, and it can be performed before every frequency sweep if needed. The drift of the zero point is smaller than the precision over extended periods of time (weeks).

**Sample light guide**

Instead of transmission lines, such as, e.g., rectangular waveguides, it is advantageous to use light pipes for broadband measurements in the millimeter-wave region. If the wavelength is significantly smaller than the diameter of the pipe, high-efficiency transmission can be achieved over relatively long distances with only weak frequency dependence [2,22,23]. This is ideal for guiding light from a source at ambient temperature to the sample in the cryostat. Our instrument uses a thin-walled, polished stainless steel tube with an inner diameter of 19 mm, which is large enough for the setup to be approximately within the limits of geometric optics, given that the largest useful wavelength is ~5 mm ($\nu \sim 60$ GHz). A copper conical condenser is used to concentrate the radiation on the sample, similar to previous work [2,24]. Specifically, we use a cone angle of 10° and a 5 mm diameter orifice, that can be masked with aluminum foil to decrease the effective diameter for smaller sample areas. Filtering of unwanted thermal and visible radiation is achieved with an ambient-temperature black polyethylene window and a cold quartz window with black polyethylene film mounted on the quartz with a small amount of Apiezon-N grease. Both windows are almost completely transparent for frequencies below 1 THz, but absorb a high fraction of the radiation at higher frequencies. The cold window is especially important, since it is critical for our absorption detection technique that a minimum of spurious thermal radiation reaches the sample. The window is a UV fused-quartz disc of 2 mm thickness, that fits snugly in a copper seat flush with the light pipe and thermalized to the outside of the probe. In order to achieve a weak and controlled thermal coupling of the sample to the helium bath (see below), the probe consists of two concentric tubes; the inner tube serves as the light guide, and the outer tube for vacuum isolation of the system from the environment.

**Direct thermal measurement of absorbed power**
Metals and superconductors reflect most of the incident light at sub-THz frequencies, with reflection coefficients sometimes higher than 99.9%. Therefore, the usual approach in infrared spectroscopy – determination of the reflected power – is unsuitable for such samples, because it would require unrealistically high resolution and extensive measurements on reference samples. At cryogenic temperatures, it is much more sensitive and convenient to determine the absorbed power directly from measurements of sample temperature changes induced by light absorption. In other words, the sample itself is used as a bolometer, whose temperature directly depends on its optical absorption coefficient at a given radiation frequency. A similar approach was used previously, e.g., for microwave spectroscopy of superconductors [13], as well as in early optical studies of superconducting gaps [14].

As in any bolometer, the design of the thermal measurement stage presents several conflicting requirements [13,25], and it is the most important and delicate aspect of the instrument. The thermal link from the sample to the helium bath must be weak to achieve high sensitivity; however, this also exacerbates issues with sensor self-heating and heating from residual thermal radiation. Moreover, to enable a chopping rate of the incident light at sufficiently high frequencies to minimize $1/f$ noise, the thermal relaxation time must be as short as possible. Hence, both the thermal conductivity and the thermal diffusivity of the weak link should not be too small. The setup is constructed such that the sample is mounted directly on a chip semiconductor temperature sensor (cernox, LakeShore Cryotronics) using a small amount of Apiezon N grease. In turn, the sensor is connected to a copper block through a plastic (Torlon) post of very low thermal conductivity (Fig. 1), whereas the block is in good thermal contact with external liquid helium.

The heat transfer between the sensor and the block occurs then predominantly through the sensor wires (42 AWG copper, about 25 mm in length), with an effective thermal conductance of approximately 50 µW/K at 1.5 K. The conductance could easily be made smaller by using a different wire material, but we have found that this value significantly reduces spurious self-heating effects, while the sensitivity remains high. If needed, the thermal contact to the bath can be further increased by the addition of a small amount of helium gas into the vacuum of the probe.

The sensor is connected as one arm of a conventional Wheatstone bridge, and the bridge imbalance is directly proportional to the change of the sample temperature. We define the static power sensitivity $\Gamma$ through

$$\frac{\Delta V}{V_0} = \Gamma P_0$$

where $\Delta V$ is the bridge imbalance voltage, $V_0$ the bridge excitation voltage, and $P_0$ the absorbed power (taken to be time-independent here). It is straightforward to show that the sensitivity is

$$\Gamma = \frac{\alpha}{R + R_0} \frac{1}{\Lambda}$$

where $R$ and $\alpha$ are the sensor resistance and its temperature coefficient, respectively, $R_0$ is the fixed reference resistance in the bridge (in our case 10 k$\Omega$), and $\Lambda$ is the thermal conductance of the weak link. $\Gamma$ can be established experimentally, without the presence of a sample, from a measurement of the dependence of $\Delta V/V_0$ on $V_0$, where the sensor self-heating power $V_0^2/R$ plays the role of $P_0$. For small values of $V_0$, the self-heating does not appreciably change the sensor temperature and $R$ can be taken to be constant, which leads to a simple linear dependence of $\Delta V/V_0$ on $P_0$ (Fig. 2). This procedure not only provides $\Gamma$, but also enables one to determine the value of $V_0$ that maximizes the signal without causing excessive self-heating. Typical values at 2 K for a CX-1050 sensor are $\Gamma = 2.5$ mW$^{-1}$ and $V_0 = 250$ mV, which yields an absolute sensitivity of ~0.6 µV/nW. With lock-in detection it is possible to obtain noise floors below 0.1 µV, yielding a detection threshold of ~100 pW and a nominal relative sensitivity for the absorbed power higher than $10^6$ at 100 GHz. Due to the increase of $\alpha$ and decrease of $\Lambda$ at lower temperatures, the sensitivity in a $^3$He system should be significantly higher. Different Cernox models or carbon-glass sensors with higher values of $\alpha$ can also increase the sensitivity at 1.5 K substantially.

The overall sensitivity of the optical absorption detection is determined by both $\Gamma$ and the thermal relaxation time of the sample/sensor system, since the incident radiation is chopped. If we assume that the limiting factor for the thermal relaxation is the conductance of the weak link, the relaxation time is

$$\tau \sim \frac{C}{\Lambda}$$

where $C$ is the combined specific heat of the sample and sensor. When the chopping frequency $\omega$ is significantly larger than the inverse relaxation time, the temperature oscillation is out of phase with the absorbed power, and the amplitude only depends on the ratio $P_0/\omega C$. This regime is commonly used in modulation calorimetry [21]. However, for our purposes the opposite, low-frequency limit is more favorable, where the temperature oscillation is in phase with the power, and the amplitude is determined by the ratio $P_0/\Lambda$. This quasi-static limit has two distinct advantages: the sample specific heat does not influence the sensitivity, so it is easy to obtain absolute absorbed power values once $\Lambda$ is known; and the sensitivity is always higher than in the high-frequency limit due to the $1/\omega$ rolloff. In practice, $1/f$ noise and drift considerations imply that the chopping frequency should be above ~1 Hz, and we typically use frequencies between 4 and 10 Hz. The samples then need to be dimensioned such that their heat capacity is sufficiently small for the setup to be in the low-frequency regime. The condition is easily verified through measurements of the phase of the temperature oscillation with respect to the chopping signal.

A secondary monitor bolometer is mounted within the condenser cone to provide a direct measurement of the incident power for normalization purposes. This is especially important for corrections of the slow drift of the source power, which can cause systematic errors during extended frequency scans. The design of the monitor bolometer follows that of ref. [2]: it is a rectangular slab cut from a 100 Ω bulk carbon resistor, with two contact wires connected using silver epoxy. The resistor is mounted on an annular Torlon holder within the cone, and its effective area is less than 10% of the cone cross-section. With a current of 10 μA, the sensor resistance is 2.2 kΩ at 4.2 K and 21 kΩ at 1.5 K, and the resistance oscillations due to light absorption are determined from sensor voltage measurements using a lock-in amplifier with AC input coupling.

**Test measurements**

In order to demonstrate the operation of the spectrometer, we have performed three sets of magnetic resonance measurements, all at a temperature of ~1.5 K: frequency-swept ferromagnetic resonance on a single crystal of $YTiO_3$ [26]; electron spin resonance on a small, sub-mg crystal of DPPH, a free-radical reference compound; and zero-field antiferromagnetic resonance on $Mn_2P_2S_6$, a layered magnetic material [27]. $YTiO_3$ is a suitable test system due to the well-defined, strong and relatively narrow resonance absorption at low temperatures, with a frequency that can be tuned in a convenient range with readily available magnetic fields. In turn, DPPH is a well-known ESR reference compound, with no magnetic ordering at 1.5 K. Finally, $Mn_2P_2S_6$ is a layered, van der Waals magnetic system with a complex excitation spectrum and strong anisotropy, which has been predicted to show two distinct resonance modes at zero external field [27]. For the $YTiO_3$ measurement we used a cryo-free superconducting magnet with variable temperature insert (Cryogenic Ltd), the DPPH experiment was performed in a resistive high-field magnet (30 T) at the LNCMI Grenoble, and we employed a 12 T high-homogeneity superconducting magnet (Oxford) for the $Mn_2P_2S_6$ measurements.

The $YTiO_3$ FMR spectrum in a field of 1.5 T applied parallel to the crystallographic *c* axis is shown in Fig. 3a. The strong resonance peak is observed slightly above 80 GHz, in agreement with previous measurements that used conventional high-frequency ESR equipment [28]. This also confirms the wavelength/frequency calibration of the laser-photomixer source. In addition to the FMR signal, we detect absorption from the sensor assembly, given that the sample is transparent except around the resonance frequency. This spurious signal is about an order of magnitude weaker than the FMR, and shows multiple maxima and minima from residual resonances in the light pipe, condenser, and photomixer optics. These are highly repeatable and can in principle be removed through calibration. We note that the background signal decreases significantly with a reflective sample of area larger than the condenser orifice, since then almost no radiation can leak past the sample and reach the sensor independently. For transparent samples, the background can be suppressed through metallization of the back surface of the sample, or by inserting thin metallic foil between sample and sensor. This also effectively doubles the optical path through the sample, which increases sensitivity.

The ESR spectra of DPPH are shown in Fig. 3b. In order to remove the background signal, the optical absorption was measured at two magnetic fields (7 T and 8 T), and we plot the ratio of the two spectra. The sharp ESR lines at frequencies corresponding to the two fields are clearly seen, and the spurious background is nearly cancelled out. This demonstrates the suitability of the instrument for magneto-optical experiments, as well as its compatibility with the noisy resistive high-field magnet environment. We note that the linewidths of the ESR lines are larger than the intrinsic DPPH linewidth and likely determined by the inhomogeneity of the magnetic field at the sample position.

Previous AFMR measurements on $Mn_2P_2S_6$ have identified two resonance modes that should be present in the zero-field limit [27]. Conventionally, ESR experiments, including FMR and AFMR, are performed by sweeping the magnetic field at a given frequency, which makes it nearly impossible to study zero-field resonances. Yet our instrument enables high-sensitivity frequency-swept ESR measurements in a broad frequency range, which we use to detect the zero-field modes directly (Fig. 3c). Their frequencies agree well with predictions from [27], and the linewidths are comparable to those previously observed in measurements at non-zero magnetic field. The spectrometer is thus complementary to field-sweep instruments, and can be used to detect spin resonances whose frequencies do not significantly depend on field.

## III.    Conclusions and summary

To conclude, we have built an optical spectrometer for sub-THz frequencies that utilizes an ultra-broadband continuous wave photomixer source and direct thermal detection of the absorbed power. The principal purpose of the instrument are optical absorption measurements in highly reflective samples at cryogenic temperatures, but a diverse range of other experiments is possible as well. We summarize here the main technical advantages and disadvantages of the design. Compared to established time-domain methods, both the sensitivity and frequency resolution are significantly higher, due to the continuous-wave nature of the spectrometer and the thermal absorption measurement technique. The instrument can therefore accommodate samples with a wide range of reflection coefficients, which is a unique advantage. Moreover, the source power peaks around 100 GHz, so the spectrometer is most efficient at frequencies where both microwave and time-domain techniques are quite limited. However, the present design operates best at low temperatures; the sensor sensitivity decreases sharply above ~10 K, while the thermal mass increases, which limits the chopping frequency. Yet with thin samples and small-mass sensors specifically designed for higher temperatures, these limitations can likely be overcome to a large degree. A perhaps more serious issue is the strongly non-monotonic frequency dependence of the effective source power, which necessitates detailed calibration for high-frequency-resolution measurements of the absolute absorbed power. The calibration can be performed using a standard metallic sample with a resistivity sufficiently high to avoid the anomalous skin effect regime; alloys like AuAg are good candidates [13]. Another feasible option might be to use the photomixer as a detector of reflected power, through sensitive measurements of the bias current in dependence on frequency. Normalization with the signal from the secondary bolometer helps to eliminate some of the source power issues, but the cancellation is not complete due to the different positions of the sample and monitor. For high-sensitivity work, it would thus be beneficial to mount the secondary sensor as close to the sample as possible. As shown by the DPPH and $Mn_2P_2S_6$ results, the instrument is ideally suited for measurements where the source power can be factored out, such as magneto-optics.

Advances in sub-THz photomixing technology enable many improvements and variations on the basic spectrometer design presented here, including the use of circularly polarized light, photomixer detection of reflected radiation, and even higher source powers [12]. Our instrument will therefore be useful for a wide range of studies, and help to address the comparative lack of instrumentation for sub-THz spectroscopy.

**Acknowledgements**

We thank Marin Lukas, Ivan Jakovac, Vladislav Kataev, Miroslav Požek and Martin Greven for comments and discussions. This work was funded by the Croatian Science Foundation, Grant No. UIP-2020-02-9494, and the Research Group Linkage Programme of the Alexander von Humboldt Foundation (Grant No. 3.4-1022249-HRV-IP), using equipment funded in part through project CeNIKS co-financed by the Croatian Government and the European Union through the European Regional Development Fund - Competitiveness and Cohesion Operational Programme (Grant No. KK.01.1.1.02.0013). SG and SH acknowledge support from the U.S. Department of Energy through the University of Minnesota Center for Quantum Materials, Grant No. DE-SC0016371. The work in Dresden was supported by the German Research Foundation through grant AL 1771/8-1 and the Dresden-Würzburg Cluster of Excellence (EXC 2147) "ct.qmat - Complexity and Topology in Quantum Matter" (project-id 390858490). We acknowledge the support of LNCMI-CNRS, a member of the European Magnetic Field Laboratory (EMFL).

**Data availability**

The data that support the findings of this study are available from the corresponding author upon reasonable request.

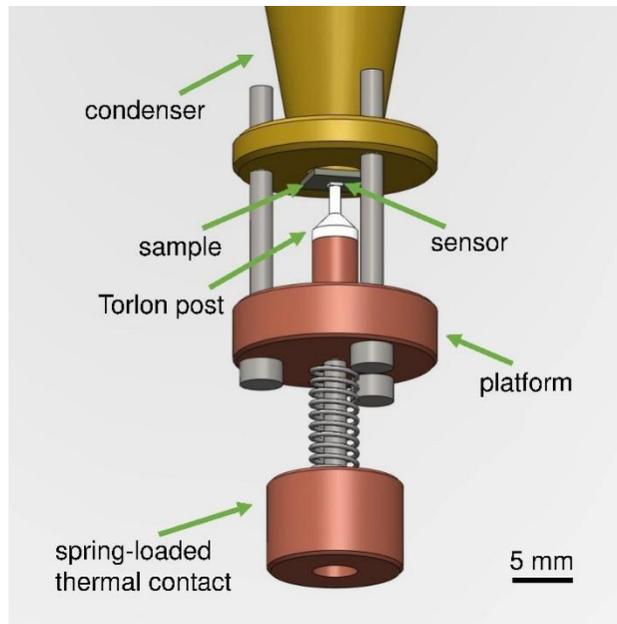

Figure 1. Schematic of the thermal power absorption measurement setup. The mm-wave light reaches the sample through a light pipe (not shown) and conical condenser. The sample is mounted on a chip temperature sensor (Cernox), which is glued to a plastic post of low thermal conductivity. A thermal link to the platform is provided by the copper sensor wires, while the platform is thermally anchored to the cryostat using a spring-loaded contact. The setup is in high vacuum, enclosed in an external stainless-steel tube with a copper bottom.

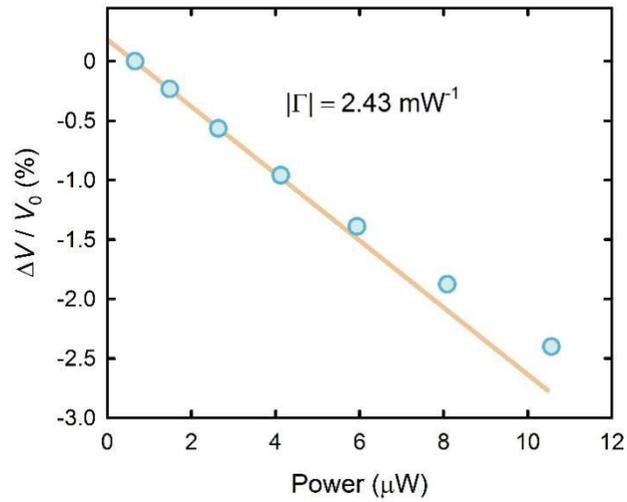

Figure 2. Calibration of the thermal sensitivity using sensor self-heating. The relative change of the bridge voltage $\Delta V/V_0$ is proportional to the self-heating power for small values of the power, with the slope equal to the sensitivity $\Gamma$. The negative sign is due to the negative temperature coefficient of resistance of the cernox sensor. The dependence on the power deviates from linear when the heating starts to induce significant temperature changes. All measured datapoints are referenced to the smallest employed excitation voltage $V_0 = 100$ mV; this is the reason for a small offset in $\Delta V/V_0$ at zero power.

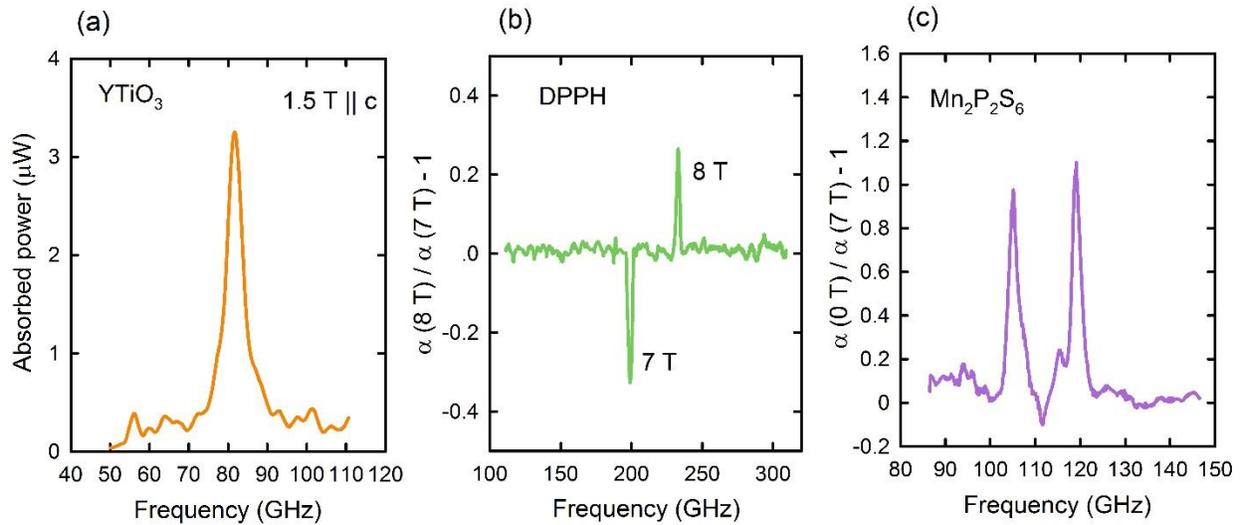

Figure 3. (a) Frequency-swept ferromagnetic resonance (FMR) measurement in a single crystal of $YTiO_3$, with an external magnetic field of 1.5 T along the crystallographic *c* direction. The background signal originates in finite absorption by the sensor assembly (the sample is transparent except around the resonance frequency); the oscillations in the intensity are repeatable and stem from standing-wave resonances in the light pipe, condenser, and photomixer. (b) Magneto-optical measurement of electron spin resonance (ESR) in DPPH, plotted as the ratio of optical absorption measurements at fields of 8 T and 7 T. This experiment was performed in a resistive high-field magnet at LNCMI Grenoble, demonstrating the compatibility of the instrument with the high-field environment. (c) Zero-field ESR spectrum of the van der Waals magnetic system $Mn_2P_2S_6$, which showcases the possibility of frequency-swept ESR measurements to detect modes that are not visible in conventional field-sweep experiments.